\documentclass[aps,prl,twocolumn,showpacs]{revtex4}

\def\beq{\begin{equation}}
\def\eeq{\end{equation}}
\def\bea{\begin{eqnarray}}
\def\eea{\end{eqnarray}}
\def\nn{\nonumber}
\def\sss{\scriptscriptstyle}
\def\bd{B_d^0}

\def\ks{K_{\sss S}}

\def\roughly#1{\mathrel{\raise.3ex\hbox
{$#1$\kern-.75em\lower1ex\hbox{$\sim$}}}}

\def\pew{P_{\sss EW}}
\def\pewc{P_{\sss EW}^{\sss C}}

\def\mpla#1#2#3{{Mod.\ Phys.\ Lett.} {\bf A#1}, #3 (#2)}

\def\plb#1#2#3{{ Phys.\ Lett.} {\bf #1B}, #3 (#2)}
\def\prd#1#2#3{{ Phys.\ Rev.} {\bf D#1}, #3 (#2)}
\def\newprd#1#2#3{{ Phys.\ Rev.} {\bf D#1}, #3 (#2)}
\def\prl#1#2#3{{ Phys.\ Rev.\ Lett.} {\bf #1}, #3 (#2)}

\def\rmp#1#2#3{{ Rev.\ Mod.\ Phys.} {\bf #1}, #3  (#2)}

\begin{document}

\preprint{UdeM-GPP-TH-03-112}
\preprint{McGill 03/20}
\title{Obtaining the Full Unitarity Triangle from $B\to \pi K$ Decays}
\author{Maxime Imbeault}
\author{Alexandre Lemerle}
\author{V\'eronique Pag\'e}
\affiliation{ Laboratoire Ren\'e J.-A. L\'evesque, 
Universit\'e de Montr\'eal, C.P. 6128, succ.~centre-ville, Montr\'eal, QC,
Canada H3C 3J7} 
\author{David London}
\affiliation{Physics Department, McGill University, 3600 University
St., Montr\'eal QC, Canada H3A 2T8; \\ Laboratoire Ren\'e
J.-A. L\'evesque, Universit\'e de Montr\'eal, C.P. 6128,
succ.~centre-ville, Montr\'eal, QC, Canada H3C 3J7}
\date{\today}
\begin{abstract}
We present a method of obtaining the entire unitarity triangle from
measurements of $B\to \pi K$ decay rates alone. Electroweak penguin
amplitudes are included, and are related to tree operators. Discrete
ambiguities are removed by comparing solutions with independent
experimental data. The theoretical uncertainty in this method is
rather small, in the range 5--10\%.
\end{abstract}

\maketitle

Within the standard model (SM), CP violation is due to a complex phase
in the Cabibbo-Kobayashi-Maskawa (CKM) quark mixing matrix. It has
become standard to parametrize this phase information using the
unitarity triangle, whose apex is given by the CKM parameters
$(\rho,\eta)$, and in which the interior (CP-violating) angles are
known as $\alpha$, $\beta$ and $\gamma$ \cite{pdg}. One of the most
important tasks in high-energy physics is to measure these quantities,
and to test whether the SM explanation is correct.

Many methods have been proposed for getting $\alpha$, $\beta$ and
$\gamma$ (or, equivalently, $\rho$ and $\eta$). Most require the
measurement of CP-violating asymmetries in hadronic $B$ decays
\cite{CPreview}. The most promising of these involve decays which are
dominated by a single decay amplitude (e.g.\ $\bd(t) \to J/\psi \ks$,
$\phi\ks$). However, a great many decays receive both tree and penguin
contributions, with different weak phases, thus spoiling the
cleanliness of the methods \cite{penguins}.

A number of years ago, it was shown that an isospin analysis of
$B\to\pi\pi$ decays allows one to remove the penguin ``pollution'' in
$\bd(t)\to\pi^+\pi^-$, so that the CP phase $\alpha$ can be measured
\cite{isospin}. Subsequently, Nir and Quinn (NQ) showed that this
technique could also be applied to $B \to \pi K$ decays, giving
another way of extracting $\alpha$ \cite{NirQuinn}. (Note that this
method gives $\alpha$ with several discrete ambiguities
\cite{ambiguities}.) Unfortunately, this analysis neglects electroweak
penguin operators (EWP's), and such operators are very important in
$B\to \pi K$ decays \cite{DeshHe}. When one includes EWP's, the NQ
$B\to \pi K$ analysis fails --- one cannot obtain weak phase
information. (The validity of analyses which rely on SU(3) relations
between $B\to \pi K$ and $B\to\pi\pi$ decays \cite{diagrams,su3} is
also affected by EWP's.)

Recently, it was shown that, by using Fierz transformations and SU(3)
symmetry, it is possible to relate EWP's to tree operators
\cite{EWPs,GPY}. In light of this, the $B\to \pi K$ analysis can be
resuscitated and improved. As we will show, the entire unitarity
triangle can be obtained from measurements of the $B\to \pi K$ decay
rates alone! In general, the discrete ambiguities can be removed by
comparison with experimental data. The theoretical error in this
method is rather small, in the range 5--10\%.

Using isospin, the $B\to \pi K$ amplitudes satisfy a quadrilateral
relation:
\beq
A^{+0} + \sqrt{2} A^{0+} = \sqrt{2} A^{00} + A^{-+} ~,
\label{quadrilateral}
\eeq
where we have defined $A^{ij} \equiv A(B \to \pi^i K^j)$. The
CP-conjugate amplitudes ${\bar A}^{ij}$ satisfy a similar relation
(note: ${\bar A}^{+0}$ corresponds to $B^- \to \pi^- {\bar K}^0$,
etc.). It is possible to express all amplitudes in terms of a number
of distinct operators. This is equivalent to a description in terms of
diagrams \cite{diagrams}. Neglecting the exchange- and
annihilation-type diagrams, which are expected to be small for
dynamical reasons, but including EWP's, there are five diagrams which
contribute to $B\to \pi K$ decays \cite{su3,su3ewp}: (1) a
color-favored tree amplitude $T$, (2) a color-suppressed tree
amplitude $C$, (3) a gluonic penguin amplitude $P$, (4) a
color-favored electroweak penguin amplitude $\pew$, and (5) a
color-suppressed electroweak penguin amplitude $\pewc$
\cite{footnote1}. The $B\to \pi K$ amplitudes can then be written
\cite{su3ewp}
\bea
A^{+0} & = & P - \frac{1}{3} \pewc ~, \nn\\
\sqrt{2} A^{0+} & = & - P - T e^{i\gamma} - C e^{i\gamma} -\pew -
\frac{2}{3} \pewc ~, \nn\\
\sqrt{2} A^{00} & = & P - C e^{i\gamma} -\pew - \frac{1}{3} \pewc ~,
\nn\\
A^{-+} & = & - P - T e^{i\gamma} - \frac{2}{3} \pewc ~.
\label{BKpiamps}
\eea
Here we have explicitly written the weak phase ($\gamma$), while $P$,
$T$, etc.\ implicitly include strong phases. To obtain the amplitudes
${\bar A}^{ij}$ for the CP-conjugate processes, one simply changes the
sign of the weak phases. We have assumed that the $b\to s$ penguin
contribution $P$ is dominated by the internal $t$-quark, so that it
has no weak phase in the Wolfenstein parametrization \cite{pdg} (the
EWP's are known to be dominated by the internal $t$-quark). In this
case the amplitudes $A^{+0}$ and ${\bar A}^{+0}$ are identical.

In the above, the (complex) $B\to \pi K$ amplitudes are written in
terms of the six complex theoretical quantities $P$, $T$, $C$, $\pew$,
$\pewc$ and $e^{i\gamma}$. First, suppose that EWP's are absent. In
this case it is possible to invert the expressions for the amplitudes
in order to write the theoretical quantities in terms of the
magnitudes and relative phases of four of the $A^{ij}$ and ${\bar
A}^{ij}$. Now, we can get the magnitudes of $A^{ij}$ and ${\bar
A}^{ij}$ from measurements of the $B \to \pi K$ branching ratios.
However, in order to obtain the relative phases, we must fix the
$A$-quadrilateral and the ${\bar A}$-quadrilateral, and know their
relative orientations. In order to do this, we need two additional
(real) relations involving the $A^{ij}$ and ${\bar A}^{ij}$. In the
absence of EWP's, such relations exist. They are:
\bea
A^{-+} + \sqrt{2} A^{00} & = & {\tilde A}^{-+} + \sqrt{2} {\tilde
A}^{00} ~, \nn\\
\sqrt{2} A^{00} + \sqrt{2} A^{0+} & = & \sqrt{2} {\tilde A}^{00} +
\sqrt{2} {\tilde A}^{0+} ~,
\label{Arelations}
\eea
where ${\tilde A}^{ij} \equiv e^{2 i \gamma} \, {\bar A}^{ij}$. (A
third relation, not independent, is $A^{+0} + A^{-+} = {\tilde A}^{+0}
+ {\tilde A}^{-+}$.) The first equation above indicates that the $A$-
and ${\tilde A}$-quadrilaterals share a common diagonal, the
isospin-3/2 amplitude:
\beq
A_{3/2} = A^{-+} + \sqrt{2} A^{00} = -(T + C)e^{i\gamma} ~.
\label{A_3/2}
\eeq
Obviously, since the diagonals are common to both quadrilaterals, they
have the same length. The second relation is used to determine this
length. With this knowledge, we can fix the $A$- and the ${\bar
A}$-quadrilaterals, and determine their relative orientations. Thus,
in the absence of EWP's, it is possible to solve for the six
theoretical quantities: $|P|$, $|T|$, $|C|$, two relative strong
phases, and $\gamma$. This is the NQ method \cite{NirQuinn}.

Unfortunately, in the presence of EWP's, it is no longer possible to
do this \cite{DeshHe,su3ewp}. In this case there are six theoretical
quantities, but only five independent amplitudes in
Eq.~(\ref{BKpiamps}). Thus, it is impossible to express the
theoretical quantities in terms of the $A^{ij}$ and ${\bar A}^{ij}$.
(It is also straightforward to verify that Eqs.~(\ref{Arelations})
above no longer hold if $\pew$ and $\pewc$ are nonzero.) It therefore
appears impossible to obtain weak phase information from $B\to \pi K$
decays.

Fortunately, to a good approximation, the EWP's are not independent
quantities --- $\pew$ and $\pewc$ can be related to $T$ and $C$.
Briefly, the argument goes as follows \cite{EWPs,GPY}. The SM
effective weak hamiltonian for $B\to \pi K$ decays is:
\beq 
H_{eff}^q = {G_F \over \protect \sqrt{2}} [V_{ub}V^*_{us}(c_1 O_1 +
c_2 O_2) - \sum_{i=3}^{10} V_{tb} V^*_{ts} c_i O_i] + h.c.
\label{H_eff}
\eeq
In the above, $O_1$ and $O_2$ are $(V-A)\times(V-A)$ tree operators,
while $O_7$--$O_{10}$ describe the electroweak penguin operators.
$O_7$ and $O_8$ have the Lorentz structure $(V-A)\times(V+A)$, while
$O_9$ and $O_{10}$ are $(V-A)\times(V-A)$. However, the Wilson
coefficients $c_7$ and $c_8$, which multiply $O_7$ and $O_8$, are much
smaller than $c_9$ and $c_{10}$ \cite{Wilson}:
\bea
c_7 = 3.49 \times 10^{-4} ~,~~ c_8 = 3.72 \times 10^{-4} ~, \nn\\
c_9 = -9.92 \times 10^{-3} ~,~~ c_{10} = 2.54 \times 10^{-3} ~.
\label{Wilson}
\eea
Thus, the EWP's are approximately given purely by $O_9$ and $O_{10}$.
Furthermore, these operators can be Fierz-transformed into $O_1$ and
$O_2$, since all have a $(V-A)\times(V-A)$ structure. Therefore the
EWP's are related to the tree operators.

There are two independent relations between EWP's and tree operators.
Ignoring exchange- and annihilation-type diagrams once again, they are
given by \cite{GPY}
\bea
P^{\sss EW}(B^+ \to \pi^+ K^0) + \sqrt{2} P^{\sss EW}(B^+ \to \pi^0
K^+) = \nn\\
{c_9 + c_{10} \over c_1 + c_2} {(T + C) \over \left\vert V_{ub}^*
V_{us} \right\vert } ~, \nn\\
P^{\sss EW}(B^0 \to \pi^- K^+) + P^{\sss EW}(B^+ \to \pi^+ K^0) = \nn\\
- {1\over 2} {c_9 - c_{10} \over c_1 - c_2} {(T - C) \over \left\vert
V_{ub}^* V_{us} \right\vert } + {1\over 2} {c_9 + c_{10} \over c_1 +
c_2} {(T + C) \over \left\vert V_{ub}^* V_{us} \right\vert } ~.
\eea
Using the expressions for the $B \to \pi K$ amplitudes given in
Eq.~(\ref{BKpiamps}), these give
\bea
\pew & = & {3\over 4} {c_9 + c_{10} \over c_1 + c_2} R (T + C) +
{3\over 4} {c_9 - c_{10} \over c_1 - c_2} R (T - C) ~, \nn\\
\pewc & = & {3\over 4} {c_9 + c_{10} \over c_1 + c_2} R (T + C) -
{3\over 4} {c_9 - c_{10} \over c_1 - c_2} R (T - C) ~,
\label{EWPtree}
\eea
where $R \equiv \left\vert V_{tb}^* V_{ts} / V_{ub}^* V_{us}
\right\vert$. These provide the relations between the diagrams $\pew$
and $\pewc$ and the tree operators $T$ and $C$.

In Refs.~\cite{EWPs,GPY}, a numerical value is taken for $R$. But in
our method, this is not necessary. Instead, we keep the general
expression
\beq
\left\vert {V_{tb}^* V_{ts} \over V_{ub}^* V_{us}} \right\vert = {1 \over
\lambda^2 \sqrt{\rho^2 + \eta^2}} ~.
\eeq
As we will see below, this allows us to improve considerably upon the
original NQ method.

The $B\to\pi K$ amplitudes are now written in terms of the five
(complex) theoretical quantities $P$, $T$, $C$, $e^{i\gamma}$ and
$\sqrt{\rho^2 + \eta^2}$. One can invert these expressions to write
the theoretical parameters in terms of five independent $A^{ij}$ and
${\bar A}^{ij}$ amplitudes \cite{footnote2}. However, as discussed
earlier, in order to determine these parameters, one needs two
additional (real) relations to fix the two quadrilaterals and their
relative orientation. There are several ways to obtain these. One is
to note that, at this stage, $e^{i\gamma}$ and $\sqrt{\rho^2 +
\eta^2}$ are simply arbitrary complex quantities, and are expressed in
terms of the $A^{ij}$ and ${\bar A}^{ij}$. However, there are physical
constraints on these parameters. They are:
\beq
\left\vert e^{i\gamma} \right\vert = 1 ~~,~~ {\rm Im}\left(
\sqrt{\rho^2 + \eta^2} \right) = 0 ~.
\label{relations}
\eeq
These provide the relations necessary to fix the relative orientations
of the two quadrilaterals.

\begin{table*}[tb]
\begin{tabular*}{\textwidth}{|l@{\extracolsep{\fill}}|l|l|l|l|l|l|l|l|l|l|}
\hline
& ~$|P|$ & ~$|T|$ & ~$|C|$ & $(\delta_{\sss P} - \delta_{\sss
T})$ & $(\delta_{\sss T} - \delta_{\sss C})$ & ~~~~$\rho$ & ~~$\eta$ &
$\sin 2\beta$ & ~~$S$ & $A_{\sss \pi K}^{indir}$ \\ \hline
~~(1)~~ & 0.96~~~~ & 0.42~~~~ & 0.33~~~~ & $-126.5^\circ$~~~~ &
$-28.0^\circ$~~~~ & $-$0.56~~~~ & 0.17~~~~ & ~~0.21~~~~ & 0.59~~~~ &
$-$0.41~~~~ \\
~~(2)~~ & 0.98 & 1.97 & 1.74 & $-21.0^\circ$ & $-149.2^\circ$ & $-$7.33 &
0.99 & ~~0.23 & 7.40 & ~~0.31 \\
~~(3)~~ & 1.0 & 0.3 & 0.05 & $-20.0^\circ$ & $-100.0^\circ$ & ~~0.18
& 0.38 & ~~0.76 & 0.42 & $-$0.80 \\
~~(4)~~ & 1.01 & 1.90 & 0.43 & $-31.9^\circ$ & $-55.0^\circ$ & $-$1.91
& 0.18 & ~~0.12 & 1.91 & $-$0.09 \\
~~(5)~~ & 1.02 & 1.70 & 0.55 & $-46.7^\circ$ & $-26.7^\circ$ & $-$0.96
& 0.07 & ~~0.07 & 0.96 & $-$0.03 \\
~~(6)~~ & 1.02 & 1.60 & 0.23 & $-4.6^\circ$ & $-8.6^\circ$ & $-$0.88
& 0.91 & ~~0.78 & 1.27 & $-$0.48 \\
~~(7)~~ & 1.08 & 2.05 & 0.59 & $-5.1^\circ$ & $-2.4^\circ$ &
$-$0.68 & 0.37 & ~~0.42 & 0.77 & ~~0.24 \\
~~(8)~~ & 1.38 & 3.12 & 1.18 & $-5.5^\circ$ & $-0.6^\circ$ &
$-$0.28 & 0.06 & ~~0.10 & 0.29 & $-$0.67 \\
\hline
\end{tabular*}
\caption{The 8 sets of theoretical parameters which reproduce the
``experimental data'' of Eq.~(\protect\ref{generateddata}). We also
give the predicted values of each set for $\sin 2\beta$, $S \equiv
\sqrt{\rho^2 + \eta^2}$ and $A_{\sss \pi K}^{indir}$, the indirect CP
asymmetry in $\bd(t)\to\pi^0\ks$.}
\label{solutions}
\end{table*}

Note that, in light of the relations between EWP's and tree operators,
the $A_{3/2}$ amplitude in Eq.~(\ref{A_3/2}) is given by
\beq
A_{3/2} = -(T + C) \left[ e^{i\gamma} + {3\over 2} {c_9 + c_{10} \over
c_1 + c_2} R \right] ~.
\label{A_3/2EWP}
\eeq
This is one of the diagonals of the $A$-quadrilateral. The
corresponding diagonal in the ${\bar A}$-quadrilateral is given by the
above expression, but with $\gamma \to -\gamma$. The relations in
Eq.~(\ref{relations}) imply that $|A_{3/2}| = |{\bar A}_{3/2}|$, as
was true for the case without EWP's. The magnitudes and relative
phases of the $B \to \pi K$ amplitudes are therefore obtained by
measurements of the branching ratios and the construction of the $A$-
and ${\bar A}$-quadrilaterals. This allows us to obtain all the
theoretical parameters.

The key point is that there is enough information in the $B\to \pi K$
system to extract the values of seven theoretical parameters: the
magnitudes of $P$, $T$ and $C$, two relative strong phases, and two
pieces of weak-phase information (which we take to be $\gamma$ and
$\sqrt{\rho^2 + \eta^2}$) \cite{footnote3}. Note that the knowledge of
$\gamma$ and $\sqrt{\rho^2 + \eta^2}$ is sufficient to pin down the
shape of the unitarity triangle. Thus, one can obtain the full
unitarity triangle (up to discrete ambiguities) from measurements of
the $B\to \pi K$ rates alone.

We now demonstrate numerically how the method works. Ideally, we would
use current experimental data on $B \to \pi K$ rates. Unfortunately,
although the various branching ratios have been measured, no
significant partial-rate asymmetries have yet been observed
\cite{pdg}, and our method requires at least one measurement of direct
CP violation. We therefore generate values for the ``experimental
measurements'' by assuming input values for $P$, $T$, etc. We choose
\bea
|P| = 1.0 ~,~~ \delta_{\sss P} = -18.0^\circ ~,~~ |T| = 0.3 ~,~~
\delta_{\sss T} = 2.0^\circ \nn\\
|C| = 0.05 ~,~~ \delta_{\sss C} = 102.0^\circ ~,~~ \rho = 0.18 ~,~~
\eta = 0.38 ~.
\label{inputs}
\eea
With these inputs, we find
\bea
& \left\vert A^{+0} \right\vert = \left\vert {\bar A}^{+0} \right\vert =
1.00 ~, & \nn\\
& \left\vert A^{0+} \right\vert = 0.86 ~,~~ \left\vert {\bar A}^{0+}
  \right\vert = 1.00 ~, & \nn\\
& \left\vert A^{00} \right\vert = 0.62 ~,~~ \left\vert {\bar A}^{00}
\right\vert = 0.57 ~, & \nn\\
& \left\vert A^{-+} \right\vert = 1.07 ~,~~ \left\vert {\bar A}^{-+}
\right\vert = 1.22 ~. &
\label{generateddata}
\eea
Here we have taken the values for $c_9$ and $c_{10}$ given in
Eq.~(\ref{Wilson}), along with $c_1 = 1.144$ and $c_2 = -0.308$
\cite{Wilson}.

Given this ``experimental data,'' we can solve the system for our
seven theoretical unknowns. Since the equations are nonlinear, there
will be many discretely-ambiguous solutions. For the ``data'' in
Eq.~(\ref{generateddata}), we find 16 solutions. Half of these yield
unitarity triangles which point down, i.e. $\eta < 0$. However, we
know from the kaon system that $\eta > 0$ \cite{signBK}. We therefore
exclude solutions with $\eta < 0$. The 8 remaining solutions are shown
in Table~\ref{solutions}. (Note: we have solved the system for many
different inputs [Eq.~(\ref{inputs})]. In all cases, we find either 16
or 8 solutions, half of which can be rejected because $\eta < 0$.)

Now, if the SM is correct, there are several constraints which these
putative solutions must satisfy. First, CP violation in $\bd(t) \to
J/\psi \ks$ has been measured, yielding a world average of $\sin
2\beta = 0.736 \pm 0.049$ \cite{browder}. Any solution in
Table~\ref{solutions} which does not give a value for $\sin 2\beta$ in
its $3\sigma$ range is excluded. Second, the latest 95\% c.l.\ range
for $S \equiv \sqrt{\rho^2 + \eta^2}$ is $0.356 \le S \le 0.452$
\cite{CKMfitter}. An acceptable solution must give a value for $S$ in
this range. Finally, some solutions can be eliminated by making the
mild theoretical assumption that $|P| > |T| > |C|$. (This constraint
is not essential --- we find that solutions which do not satisfy this
condition generally also violate one of the experimental constraints.)

In the particular case of Table~\ref{solutions}, the experimental
constraints alone eliminate all solutions except (3) (the true
solution). Indeed, in almost all of the cases we studied, in which we
varied the inputs in Eq.~(\ref{inputs}), we found that only a single
solution remained after imposing the constraints. Thus, it is in fact
possible to obtain the full unitarity triangle from measurements of
the $B\to \pi K$ rates alone.

Note that it is also possible to measure independently the indirect CP
asymmetry in $\bd(t) \to \pi^0 \ks$:
\beq
A_{\sss \pi K}^{indir} \equiv {{\rm Im} \left( e^{-2i\beta} A^{00*}
{\bar A}^{00} \right) \over \left\vert A^{00} \right\vert \left\vert
{\bar A}^{00} \right\vert} ~.
\eeq
The knowledge of this quantity will provide a crosscheck to the
solution(s) found above. If two solutions happen to be found, then one
can in principle distinguish between them through the measurement of
$A_{\sss \pi K}^{indir}$. And if one finds only a single solution
using the above method, $A_{\sss \pi K}^{indir}$ furnishes an
independent check of this solution.

It is possible that no solution is found which satisfies all the
constraints and independent measurements. This would then be evidence
for physics beyond the SM. Indeed, the present measurement of the
indirect CP asymmetry in $\bd(t) \to \phi\ks$ may be showing signs of
new physics: although the BaBar measurement is in agreement with the
SM prediction (within errors), the BELLE measurement disagrees at the
level of $3.5\sigma$ \cite{browder}. If this discrepancy with the SM
is confirmed, this would point specifically to new physics in the
$b\to s$ penguin amplitude \cite{LonSoni}. Since $B\to \pi K$ decays
also involve $b\to s$ penguin diagrams, they would also be affected by
this new physics. In particular, we would expect to find a discrepancy
in the values of the parameters of the unitarity triangle as extracted
using the above method and in other, independent measurements (e.g.\
$\sin 2\beta$).

Although there is some theoretical input in this method, the
uncertainty is rather small. There are three sources of theoretical
error. First, we ignore annihilation- and exchange-type diagrams,
leading to an error of $O(1\%)$ \cite{diagrams,su3ewp}.  Second, we
neglect the Wilson coefficients $c_7$ and $c_8$ compared to $c_9$ and
$c_{10}$, giving an error of about $4\%$ [see Eq.~(\ref{Wilson})].
Finally, one must take SU(3)-breaking effects into account in
Eq.~(\ref{EWPtree}). Ref.~\cite{EWPs} estimates such effects, and
finds them to be roughly $5\%$. Thus, depending on how one adds all
the uncertainties, the net theoretical error in this method is in the
range 5--10\%.

Finally, we must note that experimental errors in the $B \to \pi K$
branching ratios may make it challenging to implement this method in
practice. However, by performing a fit to all experimental data,
including constraints from $\sin 2\beta$ and $\sqrt{\rho^2 + \eta^2}$,
it should be possible to extract the full unitarity triangle.

In summary, we have presented a method of obtaining the entire
unitarity triangle from measurements of $B \to \pi K$ rates alone. It
relies on a relation between electroweak penguin amplitudes and tree
operators. One can distinguish among discretely-ambiguous solutions by
using independent experimental determinations of $\sin 2\beta$ and
$\sqrt{\rho^2 + \eta^2}$. The theoretical uncertainty is rather small,
in the range 5--10\%. At present, although $B$-factories have measured
the $B \to \pi K$ rates, no difference between the $B$ and ${\bar B}$
branching ratios has been observed yet. As soon as one observation of
direct CP violation in $B \to \pi K$ decays is made, it should be
possible to extract the full unitarity triangle using this method.


D.L. thanks M. Gronau for helpful conversations, and A. Hoecker and
L. Roos of the CKMfitter group for the latest 95\% c.l. range for
$\sqrt{\rho^2 + \eta^2}$. This work was financially supported by NSERC
of Canada.


\begin{thebibliography}{References:}
\bibitem{footnote1} In Refs.~\cite{su3,su3ewp}, the diagrams
contributing to $\Delta S = 1$ processes are written as $P'$, $T'$,
etc. In the present paper we drop the primes for simplicity of
notation.
\bibitem{footnote2} It is equivalent to work with the ${\tilde
A}^{ij}$ amplitudes instead of the ${\bar A}^{ij}$ amplitudes.
\bibitem{footnote3} Although we have assumed that $P$ is dominated by
the internal $t$-quark, there is still enough information in the
$B\to\pi K$ system to obtain all theoretical parameters using our
method if this assumption is relaxed. In this case $A^{+0} \ne {\bar
A}^{+0}$ and the individual rates for $B^+ \to \pi^+ K^0$ and $B^- \to
\pi^- {\bar K}^0$ must be measured.
\bibitem{signBK} This assumes that the sign of the bag parameter,
$B_{\sss K}$, is positive. For a discussion of this issue, as well as
the relevant references to the literature, see Y. Grossman, B. Kayser
and Y. Nir, \plb{415}{1997}{90}.
\end{thebibliography}
\end{document}